\documentclass[conference]{IEEEtran}
\IEEEoverridecommandlockouts
\usepackage{cite}
\usepackage{amsmath,amssymb,amsfonts}
\usepackage{algorithmic}
\usepackage{graphicx}
\usepackage{textcomp}
\usepackage{xcolor}
\usepackage{listings}
\usepackage{url}
\usepackage{titlesec}

\newcommand{\todo}[1]{}
\renewcommand{\todo}[1]{{\color{red} TODO: {#1}}}
\newcommand\scalemath[2]{\scalebox{#1}{\mbox{\ensuremath{\displaystyle #2}}}}

\lstdefinestyle{mystyle}{
    backgroundcolor=\color{backcolour},   
    commentstyle=\color{codegreen},
    keywordstyle=\color{magenta},
    numberstyle=\tiny\color{codegray},
    stringstyle=\color{codepurple},
    basicstyle=\ttfamily\footnotesize,
    breakatwhitespace=false,         
    breaklines=true,                 
    captionpos=b,                    
    keepspaces=true,                 
    numbers=left,                    
    numbersep=5pt,                  
    showspaces=false,                
    showstringspaces=false,
    showtabs=false,                  
    tabsize=2
}

\usepackage[T1]{fontenc}
\usepackage{listings}
\usepackage{url}
\usepackage{titlesec}
\usepackage{graphicx}
\graphicspath{ {./images/} }
\usepackage{float}

\usepackage{caption}
\usepackage{subcaption}

\def\BibTeX{{\rm B\kern-.05em{\sc i\kern-.025em b}\kern-.08em
    T\kern-.1667em\lower.7ex\hbox{E}\kern-.125emX}}
\begin{document}

\title{FFTc: An MLIR Dialect for Developing HPC Fast Fourier Transform Libraries\\

\thanks{Funding for the work is received from the European High-Performance Computing Joint Undertaking (JU), Grant Agreement No. 955811 (IO-SEA).}
}

\author{
    \IEEEauthorblockN{Yifei He, Artur Podobas, Måns I. Andersson and Stefano Markidis}
    \\\IEEEauthorblockA{KTH Royal Institute of Technology
    \\{\{yifeihe, podobas, mansande, markidis\}@kth.se}}

}

\maketitle

\begin{abstract}
Discrete Fourier Transform (DFT) libraries are one of the most critical software components for scientific computing. Inspired by FFTW, a widely used library for DFT HPC calculations, we apply compiler technologies for the development of HPC Fourier transform libraries. In this work, we introduce FFTc, a domain-specific language, based on Multi-Level Intermediate Representation (MLIR), for expressing Fourier Transform algorithms. We present the initial design, implementation, and preliminary results of FFTc.
\end{abstract}


\section{Introduction}
HPC libraries for computing Discrete Fourier Transforms (DFT) are critical computational building blocks for enabling signal processing, data analysis, and the solution of Partial Differential Equations (PDE). In particular, Fast Fourier Transform (FFT) algorithms solve DFT via $\mathcal{O} (n \log n)$ calculations, where $n$ is the input size against the naive DFT implementation corresponding to a matrix-vector multiply with complex numbers requiring $\mathcal{O}(n^2)$ calculations. 

Several algorithms for FFT have been designed, including the notorious Cooley-Tukey recursive scheme to the Stockham and Pease algorithms ~\cite{van1992computational}. FFT algorithms can be expressed using a factorized formulation, e.g., the entire FFT operation is expressed as the multiplication of matrices, and different algorithms will correspond to various factorization forms. These matrices are largely sparse, and their final computation will still rely only on $\mathcal{O} (n \log n)$ operations. Therefore, from an abstraction point of view, we can express any FFT algorithms in terms of matrix multiplications. Most importantly for this work, different factorizations are better suited than others for achieving high-performance on a given system. For instance, Stockham FFT factorization is an excellent fit for accelerators while other factorizations containing block matrices are a good fit for hierarchical memory systems. For this reason, to be capable of expressing and generating automatically and optimizing different FFT algorithms for different architectures is critical for producing high-performance FFT libraries. 

FFTW~\cite{frigo2005design} is among the most successful implementations of FFT libraries. Inspired by the FFTW design and development, in this work, we propose a new framework, called FFTc (FFT compiler), for the automatic generation of FFT algorithms using the MLIR and LLVM compiler infrastructure. To achieve this, we design a new language to express FFT algorithms using different formulations. The major contributions of this paper are the following:

\begin{itemize}
    \item We design and provide a first initial development of a domain-specific language for the automatic code generation of FFT algorithms, leveraging MLIR and LLVM infrastructure. 
    \item We collect and analyze the preliminary performance results from small-size one-dimensional FFT and compare the performance with the FFTW performance.
\end{itemize}

\section{Background}

The goal of this work is to develop a DSL for FFT calculation. A direct computation of the Fourier transform is the multiplication a DFT matrix by the input vector $x$. We can define the $DFT_N$ matrix as:  
%\begin{equation}
%w_N^{mn} = \operatorname{exp}(-2\pi i \frac{mn}{N}) \quad \text{for} \quad 0 \leq %m,n < N. 
%\end{equation}
\begin{multline*}
DFT_{{N}_{m,n}} = (\omega_N)^{mn},
\\
\quad \text{where} \quad \omega_N = \operatorname{exp}(-2\pi i/N) \quad \text{for} \quad 0 \leq m,n < N. 
\end{multline*}
%A dense matrix-vector multiplication has a complexity of $\mathcal{O}(N^2)$, so in practice, most implementations use a Fast Fourier Transform (FFT) algorithm, which has the complexity $\matchal{O}(N \operatorname{log} N)$. It is consequently also more accurate as the number of floating-point operations is much less for the FFT algorithms. 
The most famous FFT algorithm was introduced in 1965 by Cooley and Tukey. This algorithm relies on the recursive nature of DFT i.e. several small DFTs can describe a large DFT. In this paper, we use a matrix-formalism to represent FFT algorithms where a matrix-factorization of the DFT matrix into sparse and structured matrices describes each FFT algorithm. For example the Cooley-Tukey factorization of $\operatorname{DFT_4}$: 
\begin{equation}
\scalemath{0.75}{
\operatorname{DFT_{4}} = 
\underbrace{\begin{bmatrix}  1& & 1& \\
 & 1& & 1\\
 1& & -1& \\
 & 1& & -1\end{bmatrix}}_{\substack{ \operatorname{DFT_{2}} \otimes \operatorname{I_2}}}
 \begin{bmatrix}  1& & & \\
 & 1& & \\
 & & 1& \\
 & & & -i\end{bmatrix}
 \underbrace{\begin{bmatrix}  1& 1& & \\
 1& -1& & \\
 & & 1& 1\\
 & & 1& -1\end{bmatrix}}_{\substack{ \operatorname{I_2} \otimes \operatorname{DFT_{2}}}}
 \begin{bmatrix}  1& & & \\
 & & 1& \\
 & 1& & \\
 & & & 1\end{bmatrix},
 }
\end{equation} 
where the $I$ is the identity matrix. Here, we see the use of $\operatorname{DFT_2}$ in the formulation of $\operatorname{DFT_4}$. In the example, we see the sparse (zeros in the matrices are omitted for clarity) and structured nature of the algorithm. The Cooley-Tukey general-radix decimation-in-time algorithm for N inputs can be written as: 
\begin{equation} \label{eq:Recursive FFT}
\scalemath{0.85}{
\operatorname{DFT_N} = (\operatorname{DFT_K}\otimes \operatorname{I_M})\operatorname{D^N_M}(\operatorname{I_K} \otimes \operatorname{DFT_M})\operatorname{\Pi_K^N} \quad \text{with} \quad N = MK,
}
\end{equation} 
where $\operatorname{\Pi_K^N}$ is a stride permute and $\operatorname{D_M^N}$ is a diagonal matrix of \textit{twiddle}-factors. Different FFT algorithms, such as Stockham and Pease FFT can be expressed using different factorization schemes.

%\subsection{LLVM and MLIR}
% Yifei writes this
In this work, we use the LLVM (originally for Low-Level Virtual Machine) compiler infrastructure for the development of the FFT domain-specific language. LLVM is a collection of compiler and toolchain technologies: it consists of a set of modular compiler components, including the Clang front-ends, optimizer, code generator, debugger, linker, and OpenMP runtime. Particularly important for developing portable HPC code, the LLVM compiler technologies support many targets, including x86, Arm, and GPU systems \cite{1281665}.
%\\
\\
The LLVM project also includes Multi-Level Intermediate Representation (MLIR), a project aiming at supporting the building of domain-specific compilers, and combining existing compiler infrastructure together. While MLIR (and the XLA compiler) was initially developed by Google for machine learning workloads, MLIR is widely used today for the development of domain-specific languages beyond machine and deep learning.  To solve domain-specific problems, MLIR offers the infrastructure to define and introduce high-level abstractions and transforms  \cite{9370308}. The main mechanism to extend MLIR is the development of \emph{dialects} that allow defining new operations, attributes, and types. In addition, MLIR allows using multiple dialects that can be used together within one module. Examples of existing MLIR dialects are the affine, LLVM, GPU, vector, SPIR-V dialects. In this work, we design and develop an MLIR dialect to express FFT libraries.

\section{Related Work}
Several efforts exist for the development of high-performance FFT libraries. The inspiration for developing an FFT DSL is FFTW \cite{frigo1999fast}, which is the most widely used open-source FFT library. At its heart, FFTW is an FFT compiler, based on Objective Caml, to generate Directed Acyclic Graphs (DAG) of FFT algorithms and performs algebraic optimization on them. FFTW uses a planner at runtime to recursively decompose the DFT problem into sub-problems. These sub-problems are solved directly by optimized, straight-line code that is automatically generated by a special-purpose compiler, called \emph{genfft} \cite{frigo2005design}. An additional DSL for numerical kernels including FFT is SPIRAL. SPIRAL \cite{Franchetti:18} is a program generation system for linear transforms and other mathematical functions that produces HPC code in C. SPIRAL also supports FFTs \cite{franchetti2009discrete}: it applies pattern match and rewriting to generate optimal FFT formulation for different hardware, such as multicore systems. Then, SPIRAL maps the matrix formula to high-performance C code. 

\section{Methodology: A Domain-Specific Language for FFT}
%Yifei writes. Describe what we have designed and implemented. Here a diagram is useful.\\
%FFTc cinsists of a DSL for Fast Fourier Transform, a MLIR dialect to represent the operations , attributes and types of Fast Fourier Transform, a series of progressive lowering to transform the high level operations into low level close -to-hardware instructions. The FFTc is also built for extension and evolution, the current implementation can support various kinds of recursive and iterative FFT algorithms. We have plans to look into translators, optimizers and code generators at different levels of abstraction for better performance and functionality. The execution model and compilation pipeline is shown in figure 1.` 
This section describes FFTc-- a custom Domain-Specific Language (DSL) for describing Fast Fourier Transforms (FFT). Our aspiration with FFTc is to increase the productivity of algorithm developers without any loss in performance while at the same time being able to target multiple different backends (CPUs/GPUs/etc.) with the same input source code. In short, FFTc aims to increase \textit{productivity}, \textit{portability}, and (hopefully) \textit{performance}.

\begin{figure*}[t]
    \centering
    \includegraphics[width=0.7\linewidth]{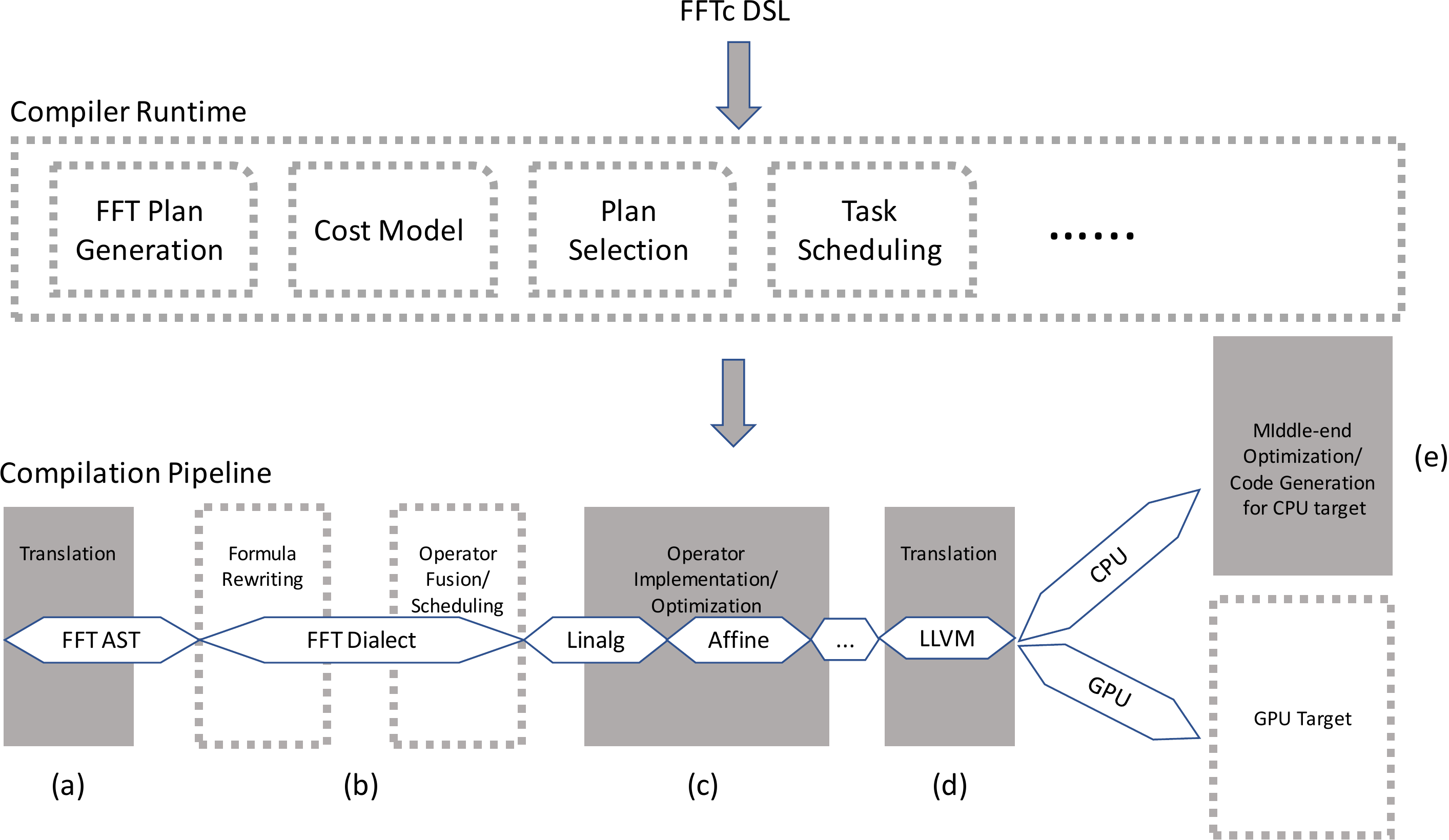}
    \caption{Compilation Pipeline}
    \label{fig:pipeline}
\end{figure*}

The execution model and compilation pipeline are shown in Figure~\ref{fig:pipeline}. The current implementation supports the parts in dark color; the remaining parts will be the focus in the near future.

The FFTc compilation pipeline has five core parts: (a) is the translation from the DSL to the Abstract Syntax Tree (AST), (b) is generating the MLIR out of AST, (c) stands for progressive lowering from FFT dialect to LLVM dialect, going through different levels of abstraction represented by dialects, (d) emits LLVM IR out of the MLIR's LLVM dialect, (e) is the LLVM middle-end compilation and code generation. 

\subsection{The FFTc language and Grammar}
%We introduce a domain-specific declarative language for FFT, which can increase the productivity for algorithm developers and reduce the effort to write high performance program for different hardware targets. 
%The notations used in the DSL is almost the same in mathematics, makes it a lot easier for the scientists to implement different kinds of FFT algorithms.
%There are domain specific information embed in the DSL, which can be passed down to compiler and reserve in the IR during the compilation pipeline, thus makes it a lot easier for the analysis and transform passes to apply aggressive optimizations.
%As shown in \ref{lst:FFTc Language}, the expressions is similar to math formula. As for binary operations, the Kronecker product is represented by '{$\otimes$}' and '{$\cdot$}' for matrix production.\\  
%The front-end is built upon the parser and lexer in LLVM's Kaleidoscope Language tutorial\cite{llvmparser}, we added support for negative numbers and the FFT related operators we defined, change the association of matrix multiplication to right associative, remain the other binary operation as left associative. A sub set of the grammar we use in Backus–Naur form is shown in \ref{lst:FFTc BNF}. 

The goal of FFTc is to create an input language that resembles (as close as possible) that of mathematics, which we believe will help end-users in being more productive without losing familiarity with the code they are writing.
An example source code of our is seen in Listing~\ref{lst:FFTc Language}, where we have aimed to keep them as similar to abstract mathematical expressions as possible, such as equation~(\ref{eq:Recursive FFT}). We support the \textit{Kronecker product} through the binary operation  '{$\otimes$}', the matrix-matrix multiplication using '{$\cdot$}', and the matrix multiplication with the twiddle matrix through the \texttt{twiddle}. Furthermore, we have a set of unary operations, such as creating the \texttt{identity} matrix, and calculating the \texttt{dft}. Finally, we have support for \texttt{permuting}. In short, we currently support all necessary language constructs to describe FFTs in a factorized form. Additionally, our grammar supports the correct right-associative binding of (e.g.,) matrix multiplication, which is different from the traditional left-associative binding of binary operators. A subset of the grammatical construct (in Backus-Naur form) is shown in Listing \ref{lst:FFTc BNF}. The grammatical construct is based on (and extended) from LLVM's Kaleidoscope language tutorial~\cite{llvmparser}.
\\
\\

\begin{lstlisting}[float=*, extendedchars=true, captionpos=b, caption= DSL FFT language, label={lst:FFTc Language}, literate={·}{{$\cdot$}}1 {⊗}{{$\otimes$}}1]
        var InputReal <4, 1> = [[1], [2], [3], [4]];
        var InputImg  <4, 1> = [[1], [2], [3], [4]];
        var InputComplex = createComplex(InputReal, InputImg);
        var result =  (DFT(2) ⊗ I(2)) · twiddle(4,2) ·
                      (I(2) ⊗ DFT(2)) · Permute(4,2) · InputComplex;
\end{lstlisting}

\begin{lstlisting}[float=*, extendedchars=true, captionpos=b, caption= {FFTc Language Grammar Extension in Backus–Naur form},label={lst:FFTc BNF}, literate={·}{{$\cdot$}}1 {⊗}{{$\otimes$}}1]
    expression -> additive-expr ('+'| '-') additive-expr
    additive-expr -> (multiplicative-expr ( '*' | '/' ) multiplicative-expr)
    multiplicative-expr -> (FFT-expr ( '*' | '/' ) FFT-expr) *
    FFT-expr -> (primary ( '⊗' | '·' ) FFT-expr) *
    primary -> identifierexpr | numberexpr | parenexpr | tensorliteral
\end{lstlisting}

\subsection{FFTc Compilation Pipeline}
The FFTc compilation pipeline shown in Figure~\ref{fig:pipeline} is based on the MLIR’s tutorial project~\cite{MLIRTutorial}. The compilation starts at the frontend (Figure~\ref{fig:pipeline}:a), where the lexical analysis, parsing, and building of an Abstract Syntax Tree (AST) based on our custom DSL language take place. The FFT dialect is the first state of MLIR generated from the AST (Figure~\ref{fig:pipeline}:b). Then a series of lowering passes are applied (Figure~\ref{fig:pipeline}:c) on the FFT dialect in order to expand many of the custom operators (e.g., the Kronecker product) into a lowered state. For example, a matrix multiplication, written in our language using "·", will be expanded to a three-level nested loop implementing said matrix multiplication. Furthermore, we can apply several existing MLIR optimization passes (such as Affine) in order to further optimize the transformed kernels. Finally, near the end of the pipeline (Figure~\ref{fig:pipeline}:c), we lower our representation to the LLVM Intermediate Representation (IR), after which we inject the code into the LLVM backend for compilation towards machine code (Figure~\ref{fig:pipeline}:d). We explain this pipeline in more detail next.

%In the future, we intend to augment our compilation pipeline with multiple phases not yet implemented (shown in Figure~\ref{fig:pipeline} as white areas), such as formula rewriting (occuring in the abstract phases in our pipeline), planning (which is available in other FFT frameworks such as FFTW[REF]), and different parallel backend (e.g., OpenMP tasks [REF]) that will be driven by some architecture-specific cost model.

%We build our compilation pipeline based on the MLIR's tutorial project \cite{MLIRTutorial}. We expand the parser to support the notations we added for the DSL. We describes the compilation pipeline in two main parts: 1) the FFT dialect we defined to represent the FFT related operations and complex data type. 2) the aggressive lowering to MLIR dialect and code generation algorithms for FFT related operations

\subsubsection{Phase 1: Translation}
\begin{figure*}[h]
    \centering
    \includegraphics[width=0.8\textwidth]{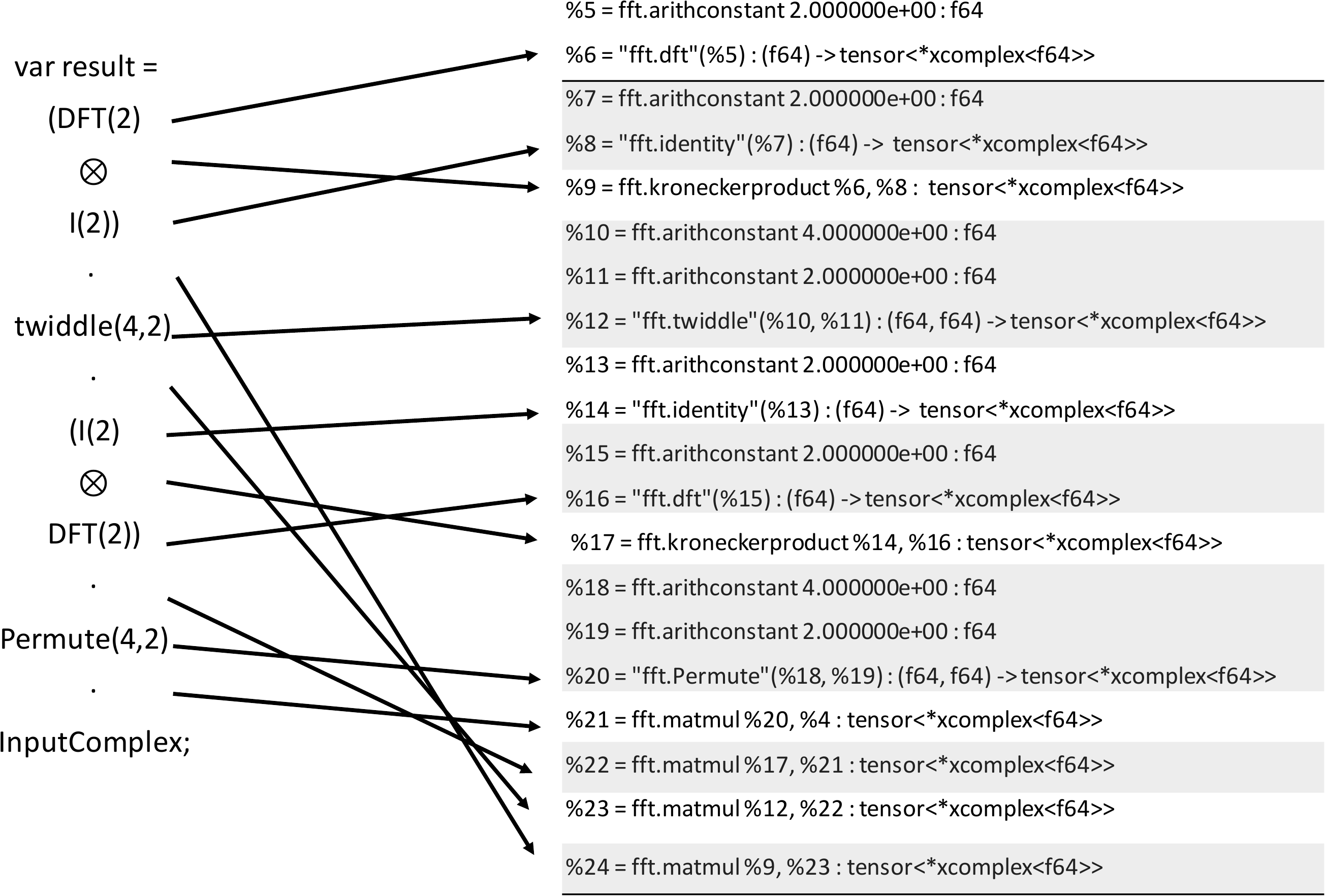}
    \caption{Mapping from the Recursive FFT to MLIR.}
    \label{fig:MapToAffine}
\end{figure*}

The FFT dialect is the first dialect in the compilation pipeline. The FFT dialect provides the basic building blocks for different kinds of FFT algorithms and defines the complex tensor data type and operations.
\begin{itemize}
\item \textbf{FFT dialect data type:} The FFT dialect operates on the double tensor and complex tensor as well as scalar integer as attributes. There is \texttt{createComplex} to generate the complex tensor from the double tensor of real and imaginary parts.
\item \textbf{FFT dialect operations:} We define the operations needed to implement various kinds of popular FFTs. Examples of such operators are the Kronecker product and matrix-matrix multiplication. We also define the \texttt{DF}, \texttt{Identity}, and \texttt{permute} matrix generator. These make it a lot easier to construct the FFT algorithm with the similar notation and syntax in mathematics.  The map of operations from FFTc DSL to MLIR FFT dialect is shown in Table~\ref{table:mapDSL2FFT}.
\end{itemize}
\begin{table}
\centering
\begin{tabular}{c c}
\textbf{\large FFTc DSL}  & \textbf{\large FFT Dialect}  \\
        \hline
createComplex(A, B) &   fft.createCT(a,b)  \\
A {$\cdot$} B & fft.matmul a, b :\\
A {$\otimes$} B & fft.kroneckerproduct a, b \\
twiddle (a,b) &  fft.twiddle (a , b) \\
I(size) & fft.identity (a) \\
DFT(size) &  fft.dft(a) \\
Permute (a ,b) &  fft.Permute(a, b) \\ [2ex]
\end{tabular}
\caption{From FFTc DSL to FFT Dialect MLIR}
\label{table:mapDSL2FFT}
\begin{tabular}{c c}
\end{tabular}
\end{table}
With the FFT dialect implementation described above, we can generate MLIR out of AST, as shown in (a) to (b) in Fig.~\ref{fig:pipeline}. Fig.~\ref{fig:MapToAffine} shows a example of the FFT dialect IR that is translated from size 4 recursive FFT in Listing \ref{lst:FFTc Language}. 

\subsubsection{Phase 2: Operator Implementation/Optimization}%\\
MLIR supports different levels of abstraction through dialects. We lower the FFT dialect to a mix of dialects. Then, we can reuse the analysis/transform passes embedded in those dialects. We run shape inference to prepare for later transforms and perform progressive lowering to a mix of dialects to implement and optimize FFT operations.
\begin{itemize}
\item \textbf{Shape Inference:} In the FFTc DSL, all the operations operate on generic tensors. We do not need to explicitly specify the shape of tensor data. This reduces the efforts of the programmers. However, carrying shape information in the IR can simplify the workload of analysis and transform passes, as well as code generation. We can obtain the shape of input tensors during the initialization of constants. Later, we propagate the shapes through the computation to every operation involved. We implement a specific shape inference function for each operation based on the input augments, such as for the Kronecker product. All dimensions of the output tensor would be the multiplication of the corresponding dimensions of two input tensors.   
\item \textbf{Progressive Lowering:} The compilation pipeline generates the actual implementations of the operations, which we defined through progressive lowering. To reuse existing optimizations in MLIR's dialects, we lower the FFT dialect to a mix of dialects, comprising of Affine, Arithmetic, Complex and MemRef dialects. The Affine dialect uses techniques from polyhedral compilation to provide a powerful abstraction for affine operations and analyses, such as dependence analysis and loop transformations. The Arithmetic dialect is intended to hold basic integer and floating-point mathematical operations, and the Complex dialect is intended to hold complex numbers creation and arithmetic operations. The MemRef dialect is intended to hold core memref creation and manipulation operations~\cite{MLIRDialects}. 

\item \textbf{Affine Dialect:} We implement the computation-heavy part of the DSL in Affine dialect, by lowering from the tensor type that FFT dialect operates on to the MemRef type that is indexed via an affine loop-nest. Tensors represent an abstract value-typed sequence of data. By using tensor and tensor operations, we can increase the productivity of algorithm developers since it is similar to the notations used in mathematics. The MemRefs dialect, on the other hand, represents the lower level buffer access, builds a bridge to the actual computer memory.

To implement the operators, we allocate a chunk of memory for the output tensor, construct loops to compute each element of the output tensor, then store them to the corresponding index of the output memory. The scalarized tensor arithmetic operations are performed by corresponding operations in the Complex dialect. The lowering result of a matrix multiplication operator is shown in the Listing~\ref{lst:Affine MatMul}. We take advantage of the existing optimizations in the Affine dialect, such as loop fusion, AffineScalarReplacement and AffineLoopInvariantCodeMotion. These optimization passes can help perform operator fusion, eliminate redundant load/store and hoists loop invariant operations out of Affine loops.
\end{itemize}
\begin{lstlisting}[float=*, extendedchars=true, captionpos=b, caption= Affine Code Example for FFT.MatMul Operation, label={lst:Affine MatMul}, literate={·}{{$\cdot$}}1 {⊗}{{$\otimes$}}1]
    From:
        %10 = fft.matmul %9, %3 : (tensor<4x4xcomplex<f64>>,
        tensor<4x1xcomplex<f64>>) -> tensor<4x1xcomplex<f64>>
    To:
        affine.for %arg0 = 0 to 4 {
        affine.for %arg1 = 0 to 1 {
            affine.for %arg2 = 0 to 4 {
            %18 = affine.load %9[%arg0, %arg2] : memref<4x4xcomplex<f64>>
            %19 = affine.load %3[%arg2, %arg1] : memref<4x1xcomplex<f64>>
            %20 = complex.mul %18, %19 : complex<f64>
            %21 = affine.load %2[%arg0, %arg1] : memref<4x1xcomplex<f64>>
            %22 = complex.add %21, %20 : complex<f64>
            affine.store %22, %2[%arg0, %arg1] : memref<4x1xcomplex<f64>>
            }
        }
    }

\end{lstlisting}
\subsubsection{Phase 3: Translation}
There exist infrastructures in MLIR to perform a full conversion from the Affine, MemRef, and Complex dialects to the LLVM dialect. Then, we can emit the LLVM IR from the LLVM dialect.
\subsubsection{Phase 4: Code Generation}
 We set up a JIT compiler using the MLIR wrapper over LLVM OrcJit, and pass the optimization and debug flags to the JIT compiler. The pass manager is also populated by MLIR. Then, the JIT compiler will perform the LLVM's middle-end optimization and code generation.

\begin{figure}[h]
    \centering
    \includegraphics[width=0.45\textwidth]{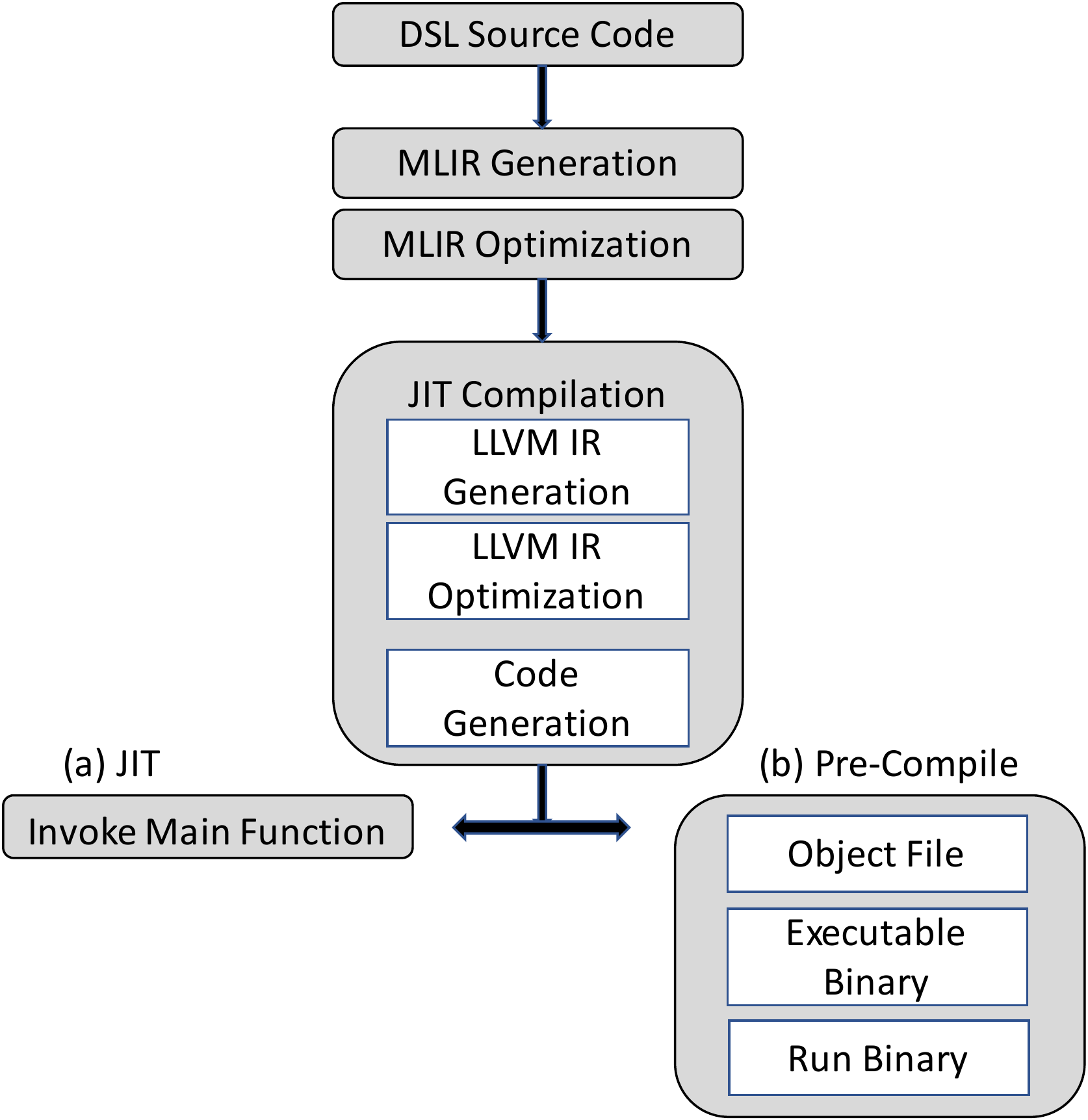}
    \caption{Compilation modes.}
    \label{fig:comp_modes}
\end{figure}

\paragraph{Ahead-Of-Time vs Just-In-Time Compilation}
%There are two compilation modes in FFTc, JIT compilation and Pre-Compilation, as shown in figure 5. The two compilation modes go through the same pipeline for generating the MLIR and perform the MLIR transforms, then generate LLVM IR from the LLVM dialect, run middle-end optimizations on LLVM IR, after that perform the code generation. \\
%The divergence comes after code generation, LLVM JIT is able to generate native code directly in memory, we can cast the result pointer to a function pointer and call it directly, there is no difference between JIT compiled code and native machine code that is statically linked. JIT mode invoke the main function and finish the compilation on the fly. 
%For the pre-compiled mode,instead of run the program immediately, we dump the object file generated in the JIT compilation, then link to executable binary and save it as independent file. So that in the run time, we can just execute the compiled binary.\\
%The two compilation modes are inline with the OpenCL's Online/Offline compilation\cite{IntelOffline}. Using the pre-compiled mode can save the time of compilation, speeding up the program significantly. In our current implementation, the FFT size needs to be constant in the compilation pipeline, that means we need to know the size of FFT if we want to use the Pre-Compiled mode. If the FFT size is a variable at runtime, we need to use the JIT mode instead.
We support two types of compilation modes in FFTc: Ahead-of-Time (AOT) and Just-in-Time (JIT) compilation. The compilation modes can be seen in Figure~\ref{fig:comp_modes}, where they share multiple components and are in line with similar compilation flows (e.g., in OpenCL's Online/Offline compilation\cite{IntelOffline}). In short, both modes start by parsing the DSL source code and transforming/optimizing it using our MLIR intermediate representation. Next, we lower the MLIR down to LLVM IR. Once in LLVM IR, the two modes differ: using the JIT mode, we directly execute the main function of our compiled targets and exit afterward. The AOT mode, instead, transforms the LLVM IR representation to an object file, links with eventual standard libraries, and outputs a machine code binary file that can be invoked by the user.\\ 
Using either model has benefits and limitations. For example, the AOT mode can be faster and speed up the final execution significantly but has the limitation that the FFT size needs to be constant. The JIT model, on the other hand, is slower but allows the FFT size to be variable at runtime. In short, the AOT mode trades flexibility for performance, while the JIT mode honors flexibility over performance.

\section{Experimental Setup}
% Yifei writes

We evaluate our FFTc on the Kebnekaise supercomputer that is located at the HPC2N HPC center in Umeå, Sweden. Kebnekaise nodes have a dual-socket Intel Xeon Gold 6132 CPU, 192 GB of RAM. The operating system is Ubuntu 20.04.4 LTS. The version of LLVM we use to embed the FFTc is 15.0.0. We run the Ahead-of-Time compilation mode FFTc 1,000 times, and we calculate error computing the standard deviation for 30 execution rounds. We developed a Python script to generate the recursive implementation of the Cooley-Tukey FFT algorithm, using our FFTc DSL. An example of the output program is shown in Listing~\ref{lst:FFTc Language}. Albeit our script can generate different FFT algorithm implementations, in this paper, we only present the results of the recursive Cooley-Tukey algorithm.

\section{Results}

%\subsection{Correctness}
As first step of our evaluation, we verify the correctness of DSL implementation.  We test different random input vectors with different sizes: the input sizes are the powers of two, from 32 to 1024. We employ complex numbers in double-precision. We compare the results with the NumPy's FFT function, that is based on FFTW. The error is calculated as $\frac{|result_{DSL} - result_{Numpy}|}{FFT_{size}}$. The error is smaller than 1e-7 for each run. 
%$\abs(resultDSL - resultNumpy) / FFTSize$. The error is smaller than 1e-7. 

\begin{figure}[h]
    \centering
    \includegraphics[width=0.45\textwidth]{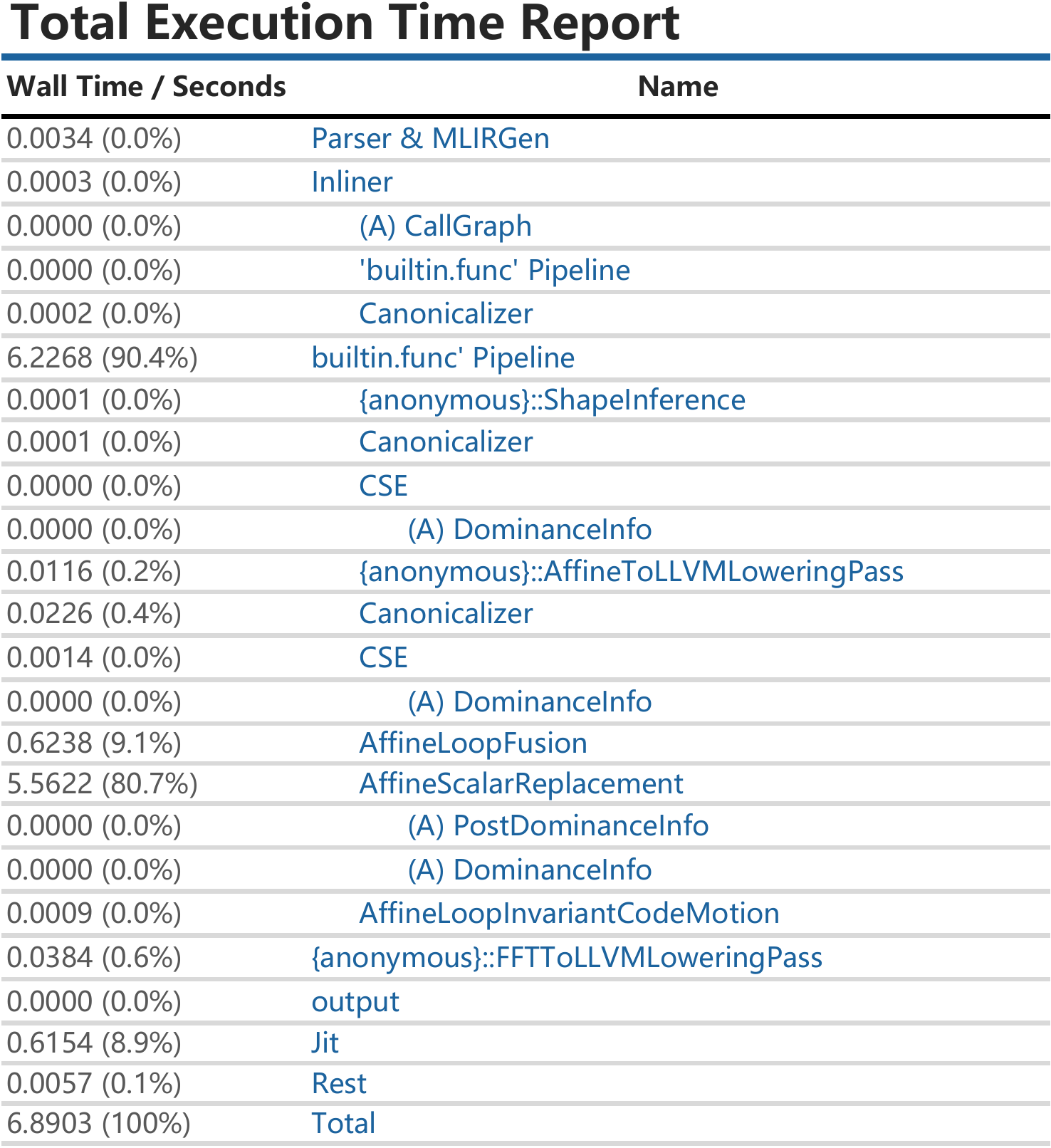}
    \caption{JIT Mode Performance for size 32 recursive FFT}
    \label{fig:jit perf}
\end{figure}

%\subsection{Performance}
%\subsubsection {JIT Mode Performance}
For the next step, we evaluate the performance in the JIT mode. We measure the execution time of size 32 recursive FFT under JIT mode. The execution time is shown in Fig.~\ref{fig:jit perf}. In the figure, the item Parser\&MLIRGen stands for frontend compilation, 'builtin.func' stands for MLIR compilation pipeline, 'Jit' stands for both LLVM Jit compilation and running time. It is clear from analyzing the figure that the frontend takes a minor portion of the execution time. The MLIR pipeline takes the largest part of the execution time. Most of the time is spent in the optimization passes such as AffineLoopFusion and AffineScalarReplacement. We can choose whether to run these optimization passes or not by passing optimization flag to FFTc, currently there are O0/O2/O3 available. The Jit part takes much smaller portion compared with MLIR pipeline, under O3 optimization option for both LLVM middle-end compilation and code generation. In actual applications, FFT algorithms may run many times while only need to be compiled once, so the compilation time does not matter considerably. As a future plan, we intend to reduce the compilation time, such as multi-threading the compiler and remove redundant operations in Affine passes. 

\begin{figure*}[h]
    \centering
    \includegraphics[width=0.75\textwidth]{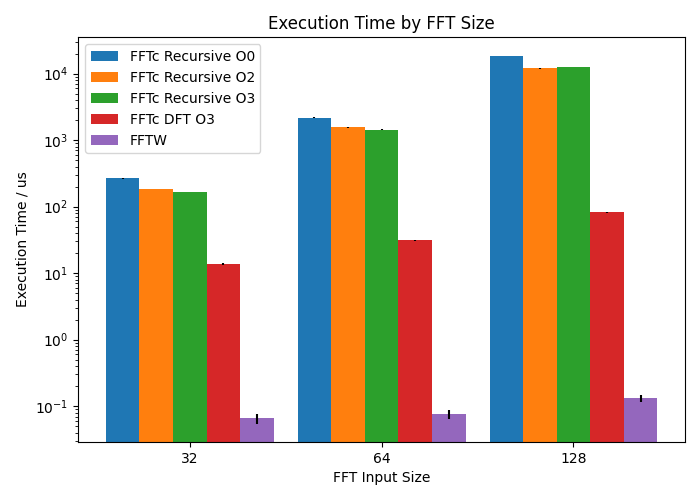}
    \caption{FFTc Single Thread Performance Compared with FFTW}
    \label{fig:FFTcwPerf}
\end{figure*}

%\subsubsection {Pre-Compiled Mode Performance}
Under Pre-Compiled mode, we compare the FFTc pre-compiled binary with FFTW 3.3. We built FFTW with gcc compiler, enabled the SIMD instructions. The input size of the FFT are the powers of 2, we use single thread to run the program. The result is shown in Fig. \ref{fig:FFTcwPerf}, the standard deviation is shown as the black lines over bars.%\\
We run four versions of FFT using FFTc: direct DFT implementation and Cooley-Tukey recursive FFT implementation with different optimization flags (O0/O2/O3). It is expected that the DFT performs much better than recursive implementations, because current implementation for FFT is computed through dense matrix multiplication, and to achieve the O(N log N) complexity FFT must be sparse matrix computation. The workload of the currently developed recursive FFT is much larger than DFT. However, we intend to use the current solution to showcase the functionality of FFTc and are planning to rewrite the computation in sparse form as a future work.
%As shown in the zoom in box(we don't have that yet),
The performance with optimization flag O3 is better that O2 and O0. The difference between O2 and O3 flag is that under O2, the AffineScalarReplacement pass will not be executed. For size 128 the O2 is slightly better than O3. Investigating the MLIR, the AffineScalarReplacement performs memory access optimizations. In addition, there is also a similar optimization pass in LLVM pipeline. We plan to further investigate this issue in the future.

When comparing the performance between FFTc Cooley-Tukey code and FFTW, we note that here is still a significant performance gap. We believe that this gap can be attributed to (amongst others) the following reasons:
\begin{itemize}
    \item The recursive factorized FFTs are computed through matrix-matrix multiplication where the matrices are not expressed as sparse matrices.
    \item We do not take full advantage of MLIR/LLVM infrastructure to generate high performance code. Examples of such a features are loop tiling, unrolling and jam and vectorization in the MLIR/LLVM pipeline.
    \item We do not support yet an autotuning mechanism, such as the FFTW \emph{planner}, to decompose the FFT problem into simpler sub-problems, later solve the simpler sub-problems using codelets generated by genfft. Currently, our implementation is similar to genfft: for the FFTs with large-size input, the generated code is extremely large and introduces considerable compilation overhead.  
\end{itemize}

\section{Discussion and Conclusion}
In this paper, we have introduced FFTc-- an emerging, work-in-progress DSL for describing different FFTs variants. The goal of FFTc is to decouple algorithm description from hardware-specific details and ultimately provide higher productivity and better portability without sacrificing performance. To this end, we have chosen an abstract language representation that is not unlike the mathematical formulas we are used to describing FFTs. We show how such an abstract language design can be mapped down-to machine code by leveraging existing MLIR and LLVM infrastructure. The performance -- while not a direct objective of this paper -- of our DSL is not yet on par with state-of-the-art FFTW, but is never-the-less a good starting point to further build upon in future performance-focused studies, such as extending our compiler with support for OpenMP tasking or vectorization.

\section{Acknowledgement}
Funding for the work is received from the European High-Performance Computing Joint Undertaking (JU), Grant Agreement No. 955811 (IO-SEA). I want to thank Steven W. D. Chien (wdchien@kth.se) for his help with the proofread.

\bibliographystyle{IEEEtran}
\bibliography{IEEEabrv, ref}

\end{document}